\documentclass[twocolumn,showpacs,preprintnumbers,amsmath,amssymb,floatfix]{revtex4}

\usepackage{amsmath}
\usepackage{epsfig}
\usepackage{graphics}
\usepackage{color}
\usepackage{graphicx}
\usepackage[font={footnotesize,it}]{caption}
\usepackage[colorlinks=true,
            linkcolor=black,
            urlcolor=blue,
            citecolor=blue]{hyperref}

\begin{document}

\title{Gravitational Lensing by Rotating Wormholes }

\author{Kimet Jusufi}
\email{kimet.jusufi@unite.edu.mk}
\affiliation{Physics Department, State University of Tetovo, Ilinden Street nn, 1200,
Tetovo, Macedonia}
\affiliation{Institute of Physics, Faculty of Natural Sciences and Mathematics, Ss. Cyril and Methodius University, Arhimedova 3, 1000 Skopje, Macedonia}

\author{Ali \"{O}vg\"{u}n}
\email{ali.ovgun@pucv.cl}
\affiliation{Instituto de F\'{\i}sica, Pontificia Universidad Cat\'olica de
Valpara\'{\i}so, Casilla 4950, Valpara\'{\i}so, Chile}

\affiliation{Physics Department, Eastern Mediterranean University
Famagusta, Northern Cyprus, Turkey}
\affiliation{TH Division, Physics Department, CERN, CH-1211 Geneva 23, Switzerland }

\date{\today }

\begin{abstract}
In this paper the deflection angle of light by a rotating Teo wormhole spacetime is calculated in the weak limit approximation.  We mainly focus on the weak deflection angle by revealing the gravitational lensing as a partially global topological effect. We apply the Gauss-Bonnet theorem (GBT) to the optical geometry osculating the Teo-Randers wormhole optical geometry to calculate the deflection angle.  Furthermore we find the same result using the standard geodesic method. We have found that the deflection angle can be written as a sum of two terms, namely the first term is proportional to the throat of the wormhole and depends entirely on the geometry, while the second term is proportional to the spin angular momentum parameter of the wormhole. A direct observation using lensing can shed light and potentially test the nature of rotating wormholes by comparing with the black holes systems.
\end{abstract}

\maketitle

\section{Introduction}

In 1935, Einstein and Rosen, proposed the existence of traversable wormholes also known as Einstein-Rosen bridges \cite{Einstein35}. Wormholes provide a shortcut through spacetime by connecting two different spacetime points. Until now, no one manage to prove them experimentally, they are only mathematical concept. Later, Wheeler showed that wormholes would be unstable and non-traversable for even a photon \cite{Wheeler55}. However, in 1988, Morris, Thorne, and Yurtsever, worked out explicitly how to convert a wormhole traversing space into one traversing time \cite{Thorne88}. Later, other types of traversable wormholes were discovered as allowable solutions to the equations of general relativity, including a variety analyzed in a 1989 paper by Matt Visser, in which a path through the wormhole can be made in which the traversing path does not pass through a region of exotic matter \cite{Visser95}.  This type of wormholes are known as thin-shell wormholes. However, exotic matter causes problem for create wormholes. Recently, it is shown that wormholes are also important to explain the quantum entanglement \cite{epr}.

In this paper, we use the solutions of the stationary and axially symmetric rotating Teo wormhole \cite{teo}. This is the first rotating wormhole solution and the most general extension of the Morris-Thorne wormhole \cite{Thorne88}. It is noted that, unfortunately, the null energy condition is violated for the rotating Teo wormhole \cite{teo}. Detection of the wormhole is another big problem. The deflection of light in Ellis wormhole \cite{Ellis} was first pointed out by Chetouani and Clement \cite{wh0}. After this the deflection of light has been investigated in a number of paper by a non-rotating wormholes. In this line of research, Tsukamoto recently has investigated the strong/weak deflection limit by the Ellis wormhole spacetime \cite{wh1,wh2,wh3}. Gravitational lensing by Ellis wormhole was also studied by Nakajima and Asada \cite{asada}. In Ref. \cite{potopov} Bhattachary and Potapov applied direct integration method, perturbation method, and invariant angle method to recover the deflection angle in Ellis spacetime. Furthermore in Ref. \cite{abe,retrolensing} the gravitational micro-lensing and  retro-lensing by Ellis wormhole has been studied. The strong limit has been studied by Dey and Sen in Janis--Newman--Winnicour and Ellis wormhole spacetimes \cite{strong1,strong2}.  Then the work by Nandi, Zhang, and Zakharov, who studied gravitational lensing in the context of a brane world model \cite{nandi}, scalar-tensor wormholes \cite{shaikh},  the wave effect in gravitational lensing \cite{yoo}, while primordial wormholes by GUTs are predicted in the early universe  \cite{nojiri}.

Almost all the work mentioned above it was devoted to non-rotating wormholes. However, from astrophysical point of view, the rotating systems are more interesting. Hence the main motivation in this paper is to explore the gravitational lensing by a rotating Teo wormhole using the GBT. In the present paper we are going to fill in this gap. This new effective method to calculate the asymptotic deflection angle provide interesting insights in the deflection of light by showing the effect of global topology. This method was recently suggested by Gibbons and Werner (GW) for static black holes  \cite{gibbons1,werner}. A new spin was put forward by Werner who extended this method to stationary black holes \cite{werner}. Then, Jusufi applied the GBT to calculate the deflection angle in the Ellis and Janis--Newman--Winnicour wormholes \cite{wh10}. More recently, using the GBT, Jusufi and Ovgun calculated the deflection angle for the quantum improved Kerr black hole pierced by a cosmic string to show the quantum effects on it \cite{kimet1,kimet2}. The deflection angle can be calculated for the charged wormholes in Einstein-Maxwell-dilaton theory using the GBT and rotating global monopole spacetime \cite{kaa,kimet3,prieslei}. Moreover, Sakalli and Ovgun showed the deflection angle of Rindler modified Schwarzschild back hole at the infra-red region \cite{aovgun}. This method also is used in various papers in Ref.\cite{asahi1,prieslei}. The main importance of this method consists in the fact that one can compute the deflection angle by integrating over a domain $S_\infty$ outside the light ray. In particular it was shown that the deflection angle can be computed via the integral \cite{gibbons1,werner}  \begin{equation} \notag
\hat{\alpha}=-\int \int_{S_\infty} \mathcal{K} \mathrm{d}A,
\end{equation}
where $\mathcal{K}$ is the Gaussian optical curvature and $\hat{\alpha}$ gives the deflection angle. Note that the above result is valid for asymptotically Euclidan optical metrics. In the case of non-asymptotically spacetimes the above equation should be modified. 

This paper is organized as follows. In section II we present the rotating Teo wormhole spacetime and then we find the Teo-Randers optical metric. We construct the optical manifold which osculates the Teo-Randes manifold using Naz{\i}m's method. In section III, we calculate the optical metric components as well as the Gaussian optical curvature. In section IV, we present the GBT and calculate the deflection angle. In section V, we derive the same result in terms of geodesics equations.  Finally we summarize our results in the conclusion section.

\section{Teo-Renders optical metric}
Let us begin by writing the famous Teo wormhole metric which describes a rotating wormhole spacetime given as follows \cite{teo}
\begin{equation} \label{1}
\mathrm{d}s^2=-N^2 \mathrm{d}t^2+\frac{\mathrm{d}r^2}{\left(1-\frac{b_0}{r}\right)}+r^2 K^2\left[\mathrm{d}\theta^2+\sin^2 \theta \left(\mathrm{d}\varphi-\omega \mathrm{d}t\right)^2\right]
\end{equation}
with 
\begin{eqnarray}
N&=&K=1+\frac{\left(4 a \cos\theta\right)^2}{r},\\
\omega &=&\frac{2a}{r^3}.
\end{eqnarray}

Note that $a$ is the spin angular momentum and $b_0$ is a positive constant with the range of the radial coordinate $r \geq b_0$. The throat of the wormhole is at $r=b_0$ with the flare-out condition \cite{naoki}
\begin{equation}
\frac{b_0-b_{0,r}r}{2 b_0^2}>0.
\end{equation}

In the case of vanishing spin angular momentum i.e. $a=0$, the Teo wormhole metric reduces to 
\begin{equation}
\mathrm{d}s^2=-N^2 \mathrm{d}t^2+\frac{\mathrm{d}r^2}{1-\frac{b_0}{r}}+r^2 K^2\left[\mathrm{d}\theta^2+\sin^2 \theta \mathrm{d}\varphi^2 \right].
\end{equation}

In what follows we shall show that the rotating Teo wormhole metric \eqref{1} gives rise to the so-called Finsler-Randers type metric of the form \cite{werner}
\begin{equation}
\mathcal{F}(x, v)=\sqrt{\alpha_{ij}(x)v^{i}v^{j}}+\beta_{i}(x)v^{i},
\end{equation}

Note that in the last equation the following condition holds $\alpha^{ij}\beta_{i}\beta_{j}<1$, with $\alpha_{ij}$ being the Riemannian metric and $\beta_{i}$ being a one-form. If we solve Eq. \eqref{1} for the null geodesic case i.e. $\mathrm{d}s^2=0$, and then reduce the problem of deflection of light in the equatorial plane by setting $\theta=\pi/2$ we find the following Teo-Randers wormhole optical metric given by
\begin{equation}\label{7}
\mathcal{F}\left(r,\varphi,\frac{\mathrm{d}r}{\mathrm{d}t},\frac{\mathrm{d}\varphi}{\mathrm{d}t}\right)=\sqrt{\Delta\left(\frac{\mathrm{d}\varphi}{\mathrm{d}t}\right)^2+\Sigma \left(\frac{\mathrm{d}r}{\mathrm{d}t}\right)^2}+\Theta \frac{\mathrm{d}\varphi}{\mathrm{d}t},
\end{equation}
in which we have used
\begin{eqnarray}\notag
\Theta &=&-\frac{r^2 \omega }{1-r^2\omega^2},\\\notag
 \Delta &=& \frac{r^2  }{(1-r^2 \omega^2)^2},\\\notag
 \Sigma &=& \frac{r }{(1-r^2 \omega^2)(r-b_0)}.
\end{eqnarray}

Note that in the equatorial plane $N=K\to 1$. The key point about the Teo-Randers optical metric $\mathcal{F}$ given by Eq. \eqref{7} relies in the fact that finding null geodesics in a stationary spacetime metric \eqref{1} is equivalent to finding the geodesics of a Teo-Randers optical metric. This can be clearly seen if we set by $\mathrm{d}t=\mathcal{F}(x,\mathrm{d}x)$. This suggest that one can study the light deflection by simply generalizing the Fermat's principle in the framework of the Rander-Finsler type metric which states that 
\begin{equation}
\delta\,\int\limits_{\gamma_\mathcal{F}}\mathcal{F}(x, \dot{x})\mathrm{d}t=0.
\end{equation}

Where $\gamma_\mathcal{F}$ is a geodesic of the Teo-Randers wormhole optical metric $\mathcal{F}$. The Randers-Finsler metric is characterized  by the Hessian 
\begin{equation} \label{9}
g_{ij}(x,v)=\frac{1}{2}\frac{\partial^{2}\mathcal{F}^{2}(x,v)}{\partial v^{i}\partial v^{j}}.
\end{equation}
where  $x\in \mathcal{M},\ v\in T_x M$. It is worth noting that ${\cal M}$ is a smooth manifold and $T_x M$ donates the tangent space of vectors $v$ at a given point \cite{{werner}}. Having found the Teo-Randers optical metric, we can continue our discussion to construct the so-called Riemannian manifold  $(\mathcal{M}, \bar{g})$ which osculates the Teo-Randers manifold $ (\mathcal{M}, \mathcal{F}) $. This can be done by applying the Naz{\i}m's method. One can do this by simply choosing a vector field $\bar{v}$ tangent to the geodesic $\gamma_{\mathcal{F}}$, such that $\bar{v}(\gamma_{\mathcal{F}})=\dot{x}$. The Hessian reads 
\begin{equation}\label{10}
\bar{g}_{ij}(x)=g_{ij}(x,\bar{v}(x)).
\end{equation}

The motivation behind this construction relies in the fact that the geodesic $\gamma_{\mathcal{F}}$ of the Randers manifold is also a geodesic $\gamma_{\bar{g}}$ of $(\mathcal{M},\bar{g})$ (see for example \cite{werner}):
\begin{equation}
\ddot{x}^i+\Gamma^{i}_{jk}(x,\dot{x})\dot{x}^{j}\dot{x}^{k}=\ddot{x}^i+{\bar{\Gamma}}^{i}_{jk}(x)\dot{x}^{j}\dot{x}^{k}=0
\end{equation}
 in other words $\gamma_{F}=\gamma_{\bar{g}}$. We shall consider a region $S_{R}\subset M$ which is bounded by the light ray $\gamma_{\mathcal{F}}$ and a curve $\gamma_{R}$. Furthermore these curves can be parameterized as follows
\begin{eqnarray}
\gamma_{\mathcal{F}} &:& \,\,\,x^i(t)=\eta^i(t),\,\,\,t \in [0,l]\\
\gamma_{R}&:&\,\,\, x^i(t)=\zeta^i(t), \,\,\,t \in [0,l^{\star}].
\end{eqnarray}

Next one can introduce $\tau=t/l $ along the geodesic $\gamma_{F}$ which belongs to $\in (0,1)$, and $\tau^{\star}=1-t/l$ $\in (0,1)$ along the curve $\gamma_{R}$ such that we can pair each point $\eta^i(\tau)$ on $\gamma_{\mathcal{F}}$ with ${\zeta^i}(\tau^{\star})$ on $\gamma_{R}$ if we let $\tau=\tau^\star$. Now by introducing a new parameter $\sigma$ which belongs to $\in (0,1)$ we can construct a family of smooth curves $x^i(\sigma,\tau)$ such that for each point pair there is precisely one curve which touches the boundary curve. At the boundary the function $x^{i}(\sigma,\tau)$ touches the curve $\gamma_{\mathcal{F}}$ as $\eta^i(\tau)=x^{i}(0,\tau)$, and hence
\begin{equation}
\dot{\eta}^{i}(\tau)=\frac{\mathrm{d}\eta^i}{\mathrm{d}t}(\tau)=\frac{\mathrm{d} x^i}{\mathrm{d} \sigma}(0, \tau).
\end{equation}

In a similar way, the function $x^{i}(\sigma,\tau)$ touches the curve $\gamma_{R}$ as $\zeta^i(\tau)=x^{i}(1,\tau)$ with
\begin{equation} \label{15}
\dot{\zeta}^{i}(\tau)=\frac{\mathrm{d}\zeta^i}{\mathrm{d}t}(\tau)=\frac{\mathrm{d} x^i}{\mathrm{d} \sigma}(1, \tau).
\end{equation}

Moreover we can define a smooth and nonzero tangent vector field of this curves given as follows
\begin{equation} \label{16}
\bar{v}^i(x(\sigma,\tau))=\frac{\mathrm{d}x^i}{\mathrm{d}\sigma} (\sigma,\tau).
\end{equation}
where  \cite{werner}
\begin{eqnarray}\notag
x^{i}(\sigma,\tau)&=&\eta^i(\tau)+\dot{\eta}^{i}(\tau)\sigma+\mathcal{A}(\tau)\sigma^2+\mathcal{B}(\tau)\sigma^3 \\
&+& y^i(\sigma,\tau)(1-\sigma)^2 \sigma^2,
\end{eqnarray}
with
\begin{eqnarray}\notag
\mathcal{A}(\tau)&=& 3\zeta^{i}(\tau)-3\eta^{i}(\tau) -\dot{\zeta}^{i}(\tau)-2 \dot{\eta}^{i}(\tau)\eta^{i}(\tau),\\\notag
\mathcal{B}(\tau)&=& 2 \eta^{i}(\tau)-2 \zeta^{i}(\tau)+\dot{\zeta}^{i}(\tau)+\dot{\eta}^{i}(\tau).
\end{eqnarray}

\begin{figure}[h!]
\center
\includegraphics[width=0.49\textwidth]{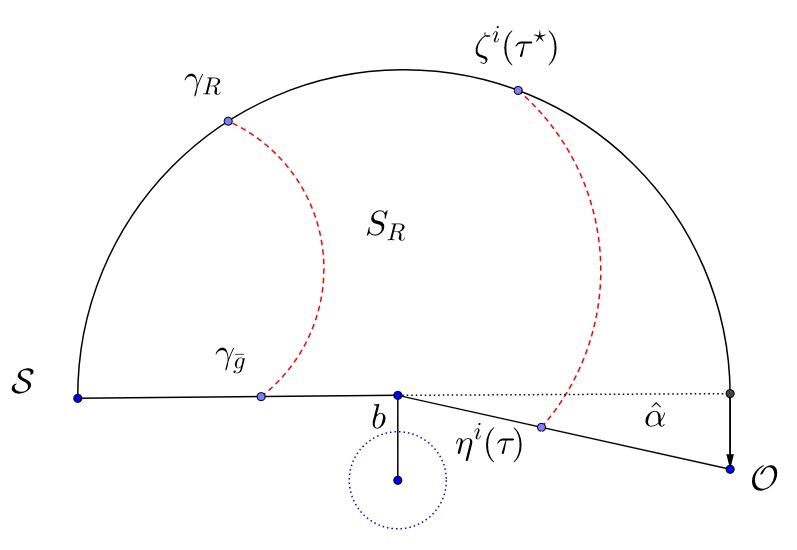} 
\caption{\small \textit {Deflection angle of light in wormhole geometry in the equatorial plane $(r,\varphi)$. Note that $b$ is the impact parameter, and $y^i(\sigma, \tau)=0$ near the light ray. In our setup we have also assumed that $b_0<<b$. }}
\end{figure}

In what follows we shall use Eqs. \eqref{16} and \eqref{10} to compute the metric components and the  Gaussian optical curvature to our osculating Riemannian geometry $(\mathcal{M},\bar{g})$ which will led us to the deflection angle using the GBT. 
Furthermore we shall calculate the deflection angle in first order terms, therefore near the light ray we can choose the undeflected light rays as 
\begin{equation}\label{17}
r(\varphi)=\frac{b}{\sin\varphi},
\end{equation}
with $b$ being the impact parameter which is approximated as the distance of the closest approach from the center of the wormhole. Making use of the Eqs. \eqref{17} and \eqref{15} one can easily convince himself that 
\begin{equation} \label{18}
\bar{v}^{r}=\frac{\mathrm{d}r}{\mathrm{d}t}=-\cos\varphi,\,\,\,\bar{v}^{\varphi}=\frac{\mathrm{d}\varphi}{\mathrm{d}t}=\frac{\sin^{2}\varphi}{b}.
\end{equation}

\section{Gaussian optical curvature}

Let us now compute the metric components. To do so, we can use Eqs. \eqref{9}, \eqref{18} for the metric components we find
\begin{eqnarray}\notag
\bar{g}_{rr}&=& \frac{r}{r-b_0}-\frac{2 a r^2 \sin^6\varphi}{(r-b_0)\left[ \frac{r\left((r-b_0)r \sin^4\varphi+b^2\cos^2\varphi \right)}{(r-b_0)} \right]^{3/2}} ,\\
&+& \mathcal{O}(a^2),\\\notag
\bar{g}_{r \varphi}&=& \frac{2 a \cos^3\varphi r}{(r-b_0)^2\left[ \frac{r\left((r-b_0)r \sin^4\varphi+b^2\cos^2\varphi \right)}{(r-b_0)b^2} \right]^{3/2}},\\
&+& \mathcal{O}(a^2),\\\notag
\bar{g}_{\varphi\varphi}&=& r^2-\frac{6 a \sin^2\varphi \left[ \frac{2 r (r-b_0)\sin^4\varphi}{3}+b^2 \cos^2\varphi \right]r}{ (r-b_0)\left[ \frac{r\left((r-b_0)r \sin^4\varphi+b^2\cos^2\varphi \right)}{(r-b_0)} \right]^{3/2}}\\
&+& \mathcal{O}(a^2),
\end{eqnarray}
with a determinant given as
\begin{eqnarray}\notag
\det \bar{g}&=&\frac{r^3}{r-b_0}-\frac{6 a r \sin^2 \varphi}{\sqrt{r\left((r-b_0)r \sin^4\varphi+b^2\cos^2\varphi \right)(r-b_0)}}\\
&+& \mathcal{O}(a^2).
\end{eqnarray}

The Gaussian optical curvature then can be found by noticing the related $\bar{R}_{r\varphi r\varphi}=\mathcal{K}\,\det \bar{g}$. In other words we can compute  $\mathcal{K}$ as follows
\begin{equation}
\mathcal{K}=\frac{1}{\sqrt{\det \bar{g}}}\left[\frac{\partial}{\partial \varphi}\left(\aleph\,\bar{\Gamma}^{\varphi}_{rr}\right)-\frac{\partial}{\partial r}\left(\aleph\,\bar{\Gamma}^{\varphi}_{r\varphi}\right)\right],
\end{equation}
where 
\begin{equation}
\aleph=\frac{\sqrt{\det \bar{g}}}{\bar{g}_{rr}}.
\end{equation}

Our computation reveals the following result 
\begin{equation}\label{23}
\mathcal{K}=-\frac{b_0}{2r^3}-\frac{3\, a }{r^{2}} \,f(r,\varphi)+\mathcal{O}(a^2),
\end{equation}

Note that the first term corresponds to the static wormhole while the second terms give the contribution of the rotation. Furthermore in the second term, for simplicity, we shall neglect the second order terms like $a b_0 \to 0 $. It is noted that the function $f(r,\varphi)$ is given by
\begin{widetext}
\begin{eqnarray}
f(r,\varphi)&=&\frac{\sin^3\varphi}{\left( {r}^{2} \sin^4\varphi+b^2\,\cos^2\varphi\right) ^{7/2}}\times \Big[ -2\,{r}^{5} \sin^{11}\varphi+10\,{b}^{2}
{r}^{3} \cos^4\varphi \sin^5\varphi\\\notag
&-& 16\,{b}^{
3}{r}^{2} \cos^4\varphi \sin^4\varphi -8\,{b}^{3}{r}^{2} \cos^2\varphi \sin^6\varphi-10\,\cos^6\varphi \sin\varphi \,{b}^{4}r- 9\,{b}^{4}r \cos^4\varphi \sin^3\varphi \\\notag
&-&  4\,{b}^{4}
r \cos^2\varphi \sin^5\varphi+4\, \cos^6\varphi {b}^{5}+ 2\,{b}^{5}\cos^4\varphi \sin^2\varphi+4\,{b}^{2}{r}^{3} \cos^2\varphi \sin^7\varphi+ {b}^{2}{r}^{3} \sin^9\varphi \Big].
\end{eqnarray}
\end{widetext}

\section{Deflection angle}
\textbf{Theorem:}
\textit{Let  $(S_{R},\bar{g})$ be a non-singular and simply connected domain over the osculating Riemannian manifold $(\mathcal{M},\bar{g})$ bounded by circular curve $\gamma_ {R}$ and the geodesic $\gamma_{\bar{g}}$. Let $\mathcal{K}$ be the Gaussian curvature of $(\mathcal{M},\bar{g})$, and $\kappa$ the geodesic curvature of $\partial S_{R}=\gamma_{\bar{g}}\cup \gamma_ {R}$. Then GBT  states that} \cite{gibbons1,werner}
\begin{equation}\label{19-4}
\iint\limits_{S_{R}}\mathcal{K}\,\mathrm{d}A+\oint\limits_{\partial S_{R}}\kappa\,\mathrm{d}t+\sum_{i}\alpha_{i}=2\pi\chi(S_{R}).
\end{equation}

Note that $\mathrm{d}A$ is the element of area of the surface, $\alpha_{i}$ are the $i^{th}$ exterior angles, while $\chi(S_{R})$ is the Euler characteristic number. In order to see how the deflection angle arises from the GBT let us  compute the geodesic curvature which gives the deviation from the geodesic. Hence if follows immediately that  $\kappa(\gamma_{\bar{g}})=0$, since $\gamma_{\bar{g}}$ is geodesic. Hence we shall calculate the geodesic curvature for the curve $\gamma_R$ as follows
\begin{equation}
\kappa (\gamma_{R})=|\nabla _{\dot{\gamma}_{R}}\dot{\gamma}_{R}|.
\end{equation}

Note that we choose $\gamma_{R}:=r(\varphi)=R=\text{const}$, with the radial part given as
\begin{equation}
\left( \nabla _{\dot{\gamma}_{R}}\dot{\gamma}_{R}\right) ^{r}=\dot{\gamma}_{R}^{\varphi
}\,\left( \partial _{\varphi }\dot{\gamma}_{R}^{r}\right) +\bar{\Gamma} _{\varphi
\varphi }^{r}\left( \dot{\gamma}_{R}^{\varphi }\right) ^{2}. \label{12}
\end{equation}

In the last equation it is  easy to notice that the first term vanishes. The second term can be calculated via the unit speed condition i.e., $\bar{g}_{\varphi \varphi} \dot{\gamma}_{R}^{\varphi } \dot{\gamma}_{R}^{\varphi }=1$. A simple calculation reveals that 
$\kappa(\gamma_{R}) \to R^{-1}$ as $R\to \infty $. Finally, letting  $R\rightarrow \infty $, the jump angles ($\alpha _{\mathcal{O}}$, $%
\alpha _{\mathcal{S}}$) gives $\pi /2$, or that is to say, the sum of jump angle to the source $\mathcal{S}$, and observer $\mathcal{O}$, yields;  $\alpha_{\mathit{O}%
}+\alpha_{\mathit{S}}\rightarrow \pi $ \cite{gibbons1}. Note that from \eqref{7} if follows
\begin{eqnarray}\notag
\lim_{R\to \infty}\mathrm{d}t &=&\lim_{R\to \infty}\left[\sqrt{\frac{R^2  }{(1-R^2 \omega^2)^2}}-\frac{R^2 \omega }{1-R^2\omega^2}\right]\mathrm{d}\varphi\\
&\to & R \mathrm{d}\varphi.
\end{eqnarray}

Note that in the last equation we have set $\omega\to 0$, since $\lim\limits_{R\to \infty} \omega=\lim\limits_{R\to \infty} \frac{2a}{R^3}\to 0$. As a result we have
\begin{equation}
\lim_{R \to \infty} \kappa(\gamma_{R})\frac{\mathrm{d}t}{\mathrm{d} \varphi}\to 1.
\end{equation}

This result clearly reveals our assumptions that our optical metric is asymptotically Euclidean. Having computed the geodesic curvature from GBT it follows
\begin{equation}
\iint\limits_{\mathcal{S}_{R}}K\,\mathrm{d}S+\oint\limits_{\gamma_{R}}\kappa \,%
\mathrm{d}t\overset{{R\rightarrow \infty }}{=}\iint\limits_{\mathcal{S}%
_{\infty }}\mathcal{K}\,\mathrm{d}A+\int\limits_{0}^{\pi +\hat{\alpha}}\mathrm{d}\varphi
=\pi
\end{equation}
 resulting with 
\begin{equation}
\hat{\alpha}=-\iint\limits_{\mathcal{S}_{\infty }}\mathcal{K}\mathrm{d}A.
\end{equation}

After substituting the Gaussian optical curvature \eqref{23} in the last equation we find the following integral
\begin{equation}\label{355}
\hat{\alpha} \simeq  -\int\limits_{0}^{\pi}\int\limits_{\frac{b}{\sin \varphi}}^{\infty}\left[-\frac{b_0}{2r^3} -\frac{3\, a }{r^{2}} \,f(r,\varphi) \right]\,\sqrt{\det \bar{g}}\,\mathrm{d}r\,\mathrm{d}\varphi.
\end{equation}

Solving the first integral we find the deflection angle for the static wormhole geometry given in terms of Elliptic type integral
\begin{eqnarray}\notag
\mathcal{I}_{1}&=&-\int\limits_{0}^{\pi}\int\limits_{\frac{b}{\sin \varphi}}^{\infty}\left( -\frac{b_0}{2r^3}\right)\sqrt{\det \bar{g}}\mathrm{d}r\mathrm{d}\varphi\\
&=& \pi- 4 \sqrt{1-z}\,\text{EllipticE}\left[\frac{\pi}{4},\frac{2z}{z-1} \right],
\end{eqnarray}
where $z=b_0/b$. Approximating the above solution we find
\begin{equation}
\mathcal{I}_{1}=\frac{b_0}{b}+\mathcal{O}\left(\frac{b_0}{b}\right)^2
\end{equation}

Next, we need to solve the second integral in Eq. \eqref{355} which is singular at $0$ and $\pi$. We simply assign a value to this integral at these singular points to find
\begin{equation}
\mathcal{I}_2= -\int\limits_{0}^{\pi}\int\limits_{\frac{b}{\sin \varphi}}^{\infty}\left[-\frac{3\, a }{r^{2}} \,f(r,\varphi) \right]\,\sqrt{\det \bar{g}}\,\mathrm{d}r\,\mathrm{d}\varphi=\pm \frac{4a}{b^2}.
\end{equation}

Thus we find the total deflection angle 
\begin{equation} \label{deflectionangle}
\hat{\alpha} = \mathcal{I}_1+\mathcal{I}_2=\frac{b_0}{b}\pm \frac{4a}{b^2}.
\end{equation}

In which the positive (resp., negative) sign is for a retrograde (resp., prograde) light ray. We note that we used a straight line approximation and it's clear that only the first order terms in $b_0$ and $a$ should be correct. But it is possible to modify the integration domain $\mathcal{S}_\infty$, which should give the correct second order terms as well. Yet another possibility is to modify the vector field \eqref{18} by including second order terms proportional in $a$ and $b_0$.

\section{Geodesic Approach}

In this section we will study the problem of calculating the deflection angle in the framework of the geodesic equations. Applying the variational principle to the Teo-wormhole metric \eqref{1} with the Lagrangian 
\begin{equation}
2\,\mathcal{L}=-\dot{t}^2+\frac{\dot{r}^2}{1-b_0/r}+r^2\left(\dot{\varphi}-\frac{2 a \dot{t} }{r^3} \right)^2=0,
\end{equation}
in which we have taken the deflection of a planar photons by letting  $\theta =\pi/2$. Next, one may define two constants of motion $l$ and $\gamma$, given  as 
\begin{eqnarray}\label{44}
p_{\varphi}&=&\frac{\partial \mathcal{L}}{\partial \dot{\varphi}}=l,\\
p_{t}&=&\frac{\partial \mathcal{L}}{\partial \dot{t}}=-\gamma.
\end{eqnarray}

From where it follows that
\begin{eqnarray}
&&r^2 \left(\dot{\varphi}-\frac{2 a \dot{t} }{r^3} \right)=l,\\
&&  \dot{t}^2+\frac{2 a}{r}\left(\dot{\varphi}-\frac{2 a \dot{t} }{r^3} \right)=\gamma .
\end{eqnarray}

Introducing a new variable $u$, related to our old radial coordinate via 
\begin{equation}
u(\varphi)=\frac{1}{r},
\end{equation}
then the following important relation can be proven
\begin{equation}\label{iden}
\frac{\dot{r}}{\dot{\varphi}}=\frac{\mathrm{d}r}{\mathrm{d}\varphi}=-\frac{1}{u^2}\frac{\mathrm{d}u}{\mathrm{d}\varphi}.
\end{equation}

Before we proceed we take $\gamma=1$, without loss of generality \cite{Boyer}. Furthermore in the weak limit it suffices to approximate the impact parameter $b$  with the distance of the closest approach 
\begin{equation}
u_{max}=\frac{1}{r_{min}}=\frac{1}{b}, 
\end{equation}
thus it follows $l=b$. Finally we are left with the differential equation, 
\begin{eqnarray}\label{42}
\frac{\left(\frac{\mathrm{d}u}{\mathrm{d}\varphi}\right)^2}{u^4(1-b_0 u)}+\frac{4a^2 \Sigma^2 }{u^2 \Theta^2 }-\frac{4 a \Sigma}{u^2 \Theta}-\frac{\Sigma^2}{u^6 \Theta^2}+\frac{1}{u^2}=0,
\end{eqnarray}
where 
\begin{eqnarray}
\Sigma (u) &=& \frac{1}{u^3}-2ab \\
\Theta (u) &=& \frac{b}{u^4}+\frac{2a}{u^3}-4a^2b.
\end{eqnarray}

Rearranging the equation \eqref{42} we find
\begin{equation}
\varphi=\int_{0}^{1/b} \frac{\mathrm{d}u}{\sqrt{\left( \frac{4 a u^2 \Sigma}{\Theta}-\frac{4a^2 u^2 \Sigma^2 }{\Theta^2 }+\frac{\Sigma^2}{u^4 \Theta^2}-u^2 \right)\left(1-b_0u \right)}}  .
\end{equation}

It is well known that one may express the solution of the differential equation \eqref{42} in the form of
\begin{equation}
\Delta \varphi =\pi+\hat{\alpha},
\end{equation}
where $\hat{\alpha}$ is the deflection angle to be calculated. In other words, we can rewrite this equation as \cite{weinberg}
\begin{equation}
\hat{\alpha}=2|\varphi_{u_{max}}-\varphi_{u_{min}}|-\pi,
\end{equation}
where
\begin{equation}
\varphi=  \int_0^{1/b}  \xi(u)  \mathrm{d}u.
\end{equation}

Note that after we expand in Taylor series around $a$ was found
\begin{eqnarray}
\xi(u)=\frac{b^3 u^2-2 au-b}{\sqrt{(b^2 u^2-1)(b_0 u-1)}\left(1-b^2u^2\right)}+\mathcal{O}(a^2).
\end{eqnarray}

Yet one can proceed to introduce new variable $y=b_0/b$ and expand in Taylor series around $y$. After we evaluate the integral the deflection angle in the weak deflection limit approximation is found to be
\begin{equation}
\hat{\alpha}\simeq \frac{b_0}{b} \pm \frac{4a}{b^2}.
\end{equation} 

As expected, this result is in complete agreement with the result found in Section IV. It is interesting to compare the rotating wormhole deflection angle with the Kerr black hole deflection angle which is given by 
\begin{equation}
\hat{\alpha}_{kerr}\simeq \frac{4 m}{b} \pm \frac{4a m}{b^2},
\end{equation} 
where $m$ is the black hole mass, and $a$ is the angular momentum parameter. Below we show graphically the bending angle as a function of the impact parameter $b$. As we can see, the bending of light is stronger in the Kerr black hole for the chosen values.

\begin{figure}[h!]
\center
\includegraphics[width=0.46\textwidth]{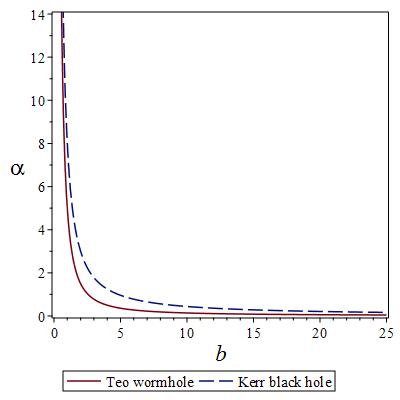} 
\caption{\small \textit {Deflection angle as a function of the impact parameter $b$. We have chosen $m=a=b_0=1$.}}
\end{figure}

\section{Conclusion}
In this paper, we have calculated the deflection angle by a rotating Teo wormhole spacetime for the first time. To our best knowledge Eq. \eqref{deflectionangle} is reported for the first time.  We have constructed the Teo-Randers optical geometry and applied the GBT to the osculating  geometry. Then we confirm our result by means of the geodesics equations. We have shown that the total deflection angle can be written as a sum of two terms. The first term depends only on the geometry and corresponds to the static wormhole case, in particular we have shown that the deflection angle is proportional to the throat of the wormhole. The second term on the other hand encodes the rotation of the wormhole and is proportional to the spin angular momentum of the wormhole. Hence, the value of the spin angular momentum $a$ affect the deflection angle. Furthermore the value of the wormhole throat $b_0$ increases the deflection angle. \\ \\

It should be noted that in the present paper we have used a straight line approximation in integrating over a  domain outside the light ray. Therefore our result is expected to be valid only in leading order terms in $b_0$ and $a$, in other words this agreement is not valid for higer-order terms. By integrating the Gaussian curvature of the optical metric outwards from the light ray, we reveal that how the global topology  plays an important role on the gravitational lensing in the wormhole spacetime. Studying weak gravitational lensing might potentially test the nature of rotating wormhole by comparing with black holes systems \cite{blackholes-wormholes1,blackholes-wormholes2,blackholes-wormholes3,sumanta}.

 \begin{acknowledgments}
We wish to thank the editor and anonymous reviewers for their valuable comments and suggestions. This work was supported by the Chilean FONDECYT Grant No. 3170035 (A\"{O}). A\"{O} is grateful to the CERN theory (CERN-TH) division for hospitality where part of this work was done.
\end{acknowledgments}

\end{document}